\documentclass{article}
\usepackage[a4paper, left=1.5cm, right=1.5cm, top=2.5cm, bottom=2.5cm]{geometry}
\usepackage{graphicx} % Required for inserting images
\usepackage{amsmath}
\usepackage{amssymb}
\usepackage{xcolor}
\usepackage{authblk}
\usepackage{cite}
\title{Is Dark Energy Inhomogeneous?}

\author[1]{Oem Trivedi\thanks{oem.trivedi@vanderbilt.edu}}
\author[2]{Abraham Loeb\thanks{aloeb@cfa.harvard.edu}}

\affil[1]{Department of Physics and Astronomy, Vanderbilt University, Nashville, TN 37235, USA}
\affil[2]{Astronomy Department, Harvard University, 60 Garden St., Cambridge, MA 02138, USA}

\date{\today}

\begin{document}

\maketitle

\begin{abstract}
The discovery of dark matter deficient galaxies motivates a fundamental question: Could sufficiently large regions of the Universe be deficient in dark energy as well? We investigate this possibility by developing a phenomenological framework that distinguishes genuine dark energy deficiencies from ordinary matter voids through nonlinear expansion, coherent peculiar velocities and density-momentum correlations. We show that a $10\%$ dark energy deficiency can suppress the expansion of a representative nonlinear void by approximately $2.7\%$, generating peculiar velocity residuals of order $60$--$100\,\mathrm{km\,s^{-1}}$ across scales of $30$--$50\,\mathrm{Mpc}$. Intriguingly, we find that deficiencies of order $20\%$ extending over several hundred megaparsecs could also influence the locally inferred Hubble constant, potentially contributing to the $H_0$ tension, subject to stringent isotropy and dynamical consistency constraints. We identify observational strategies involving DESI, Rubin, peculiar velocity surveys, and cosmic void catalogues, opening a new avenue for testing whether the dark energy driving cosmic acceleration is truly homogeneous.
\end{abstract}

\section{Introduction}
Modern cosmology has entered an era in which increasingly precise observations and rapidly developing theoretical frameworks are jointly expanding the range of physical questions that can be addressed about the Universe. A particularly interesting direction has been the study of Dark Energy (DE) and the wide variety of possible mechanisms capable of producing the observed accelerated expansion \cite{de1SupernovaSearchTeam:1998fmf,de2Li:2012dt,de3Li:2011sd,de4Mortonson:2013zfa,de5Frusciante:2019xia,de6Huterer:2017buf,de7Vagnozzi:2021quy,de8Adil:2023ara,de9Feleppa:2025clx,de10DiValentino:2020evt,de11Nojiri:2010wj,de12Nojiri:2006ri,de13Trivedi:2023zlf,de14Trivedi:2022svr,de15Trivedi:2024inb,de16Trivedi:2024dju}. While the standard $\Lambda$CDM cosmology continues to provide an exceptionally successful description of a broad range of observations, unresolved discrepancies such as the $H_0$ and $S_8$ tensions continue to motivate the exploration of extensions to the conventional picture \cite{ht1DiValentino:2021izs,ht2Clifton:2024mdy,s81kazantzidis2018evolution,s82amon2022non,s83poulin2023sigma,baoFerreira:2025lrd}. At the same time, the observational landscape is being transformed by precision measurements of the cosmic microwave background, large scale structure, distant astrophysical populations, gravitational waves and the cosmic web \cite{t1jha2019next,t2Shanks:2015lda,t3chandler2025nsf,t4Euclid:2024yrr,t5WST:2024rai,t6CosmoVerseNetwork:2025alb,t7COMPACT:2022gbl}.
\\
\\
On the other hand, Dark Matter (DM) provides an instructive example of how unusual astrophysical environments can reveal information about an otherwise poorly understood cosmic component. Although its cosmological abundance and gravitational role in structure formation are well established, its microscopic identity remains unknown \cite{dm11rubin1970rotation}. Well known candidate realizations range from weakly interacting massive particles, axions and sterile neutrinos to primordial compact objects, ultralight fields, warm dark matter and self interacting or dissipative dark sectors \cite{dm1Cirelli:2024ssz,dm2Arbey:2021gdg,dm3Balazs:2024uyj,dm4Eberhardt:2025caq,dm5Bozorgnia:2024pwk,dm6Misiaszek:2023sxe,dm7OHare:2024nmr,dm8Adhikari:2022sbh,dm9Miller:2025yyx,dm10Trivedi:2025vry,hdmTrivedi:2025sbe}. Any viable model must reproduce the observed cosmological abundance, behave approximately as a pressureless component during structure formation, possess sufficiently small sound speed and free streaming scale and remain stable over cosmological times, while simultaneously satisfying bounds on interactions with baryons, radiation and other dark sector particles. Against this broad theoretical background, the discovery of galaxies apparently containing far less dark matter than ordinarily expected has been particularly striking. NGC 1052 DF2 provided the first widely discussed example, with an internal velocity dispersion consistent with a dynamical mass close to that supplied by the stellar component alone \cite{df21van2018galaxy,df22danieli2019still,df23wasserman2018deficit}. Similar behavior was subsequently reported for NGC 1052 DF4 \cite{df41van2019second,df42montes2020galaxy,df43li2024rotation}, FCC 224 \cite{fcc1romanowsky2024candidate,fcc2tang2025unexplained,fcc3buzzo2025new} and NGC 1052 DF9 \cite{df91keim2026third,df92gannon2023keck,df93keim2025kinematic,df94van2022trail,df95tang2026new}. These diffuse systems generally exhibit unusually low internal velocity dispersions and dynamical masses comparable to their observed baryonic content, while some also possess unusual globular cluster populations or appear embedded within larger spatial and kinematic structures.
\\
\\
The growing diversity of such anomalous systems has hence shifted part of the discussion from simply asking how much dark matter a galaxy contains to asking why different cosmic components can, in some environments, apparently become spatially separated. This question becomes particularly interesting when considering very faint or "almost dark" systems such as TTT J1237327+143535 in the Virgo Cluster \cite{tttle2026serendipitous,ad1trivedi2026dark}, which can be viewed as complementary examples in which the baryonic component itself may be unusually suppressed relative to the underlying gravitating structure. Several mechanisms have been proposed to account for dark matter deficient galaxies, including strong tidal stripping, the formation of tidal dwarf galaxies from baryon rich debris and high velocity bullet dwarf collisions in which collisional gas and effectively collisionless dark matter follow different dynamical trajectories. These possibilities have also motivated the idea that such galaxies can be used not merely as astrophysical oddities, but as probes of dark matter microphysics. For example, the Dark Matter-Baryon Separability Condition \cite{trilobdm} translates the observed degree of dark matter or baryon depletion into constraints on the relative incorporation efficiencies of the two sectors. The existence of both dark matter deficient and possible baryon deficient systems suggest that spatially unusual environments may provide a general route for testing whether different cosmic components respond differently to the same formation history.
\\
\\
This prompts us to ask a very interesting question: Can there be Dark Energy deficient regions in the Universe as well, just like possible Dark Matter deficient regions that have been observed recently? If so, what could be the tests to figure out where such DE deficient regions are located, and what could these regions reveal to us about the nature of DE? Unlike DM, DE is usually treated as spatially homogeneous on astrophysical scales, with a strict cosmological constant providing the limiting case in which no propagating density perturbation exists at all. A confirmed spatial deficiency would immediately point toward a dynamical dark energy component or some more general effective gravitational sector capable of sustaining spatial structure. Such regions would also provide a new way of testing the relation between dark energy, peculiar velocities, gravitational potentials and the expansion history of underdense environments, while potentially offering an environmental contribution to existing cosmological tensions. The question would hence not only be whether dark energy can be locally deficient, but whether such regions can be defined observationally, distinguished from ordinary matter voids and connected consistently to the microphysics of cosmic acceleration. In this work, we probe this question in detail.

\section{Why consider Dark Energy deficiency?}

We discussed above how DM deficient galaxies provide a natural wonder to look towards DE deficiency too, but we would like to sharpen our motivations here. The possibility that the Universe may depart from exact homogeneity and isotropy has attracted considerable theoretical interest for decades with examples ranging from anisotropic Bianchi cosmologies to inhomogeneous Lemaitre-Tolman-Bondi (LTB) spacetimes and Kantowski-Sachs backgrounds, within which dark energy or cosmological constant contributions have also been investigated \cite{deb,deltb,deks}. Furthermore, theoretical constructions involving spatially inhomogeneous dark energy, including models in which dark matter and dark energy possess coupled or distinct perturbations, have already demonstrated that such possibilities are not without physical motivation \cite{inde1,inde2,inde3}. Particular interest has been on the influence of these departures on the evolution of gravitational potentials and the integrated Sachs-Wolfe effect, through which the spatial behavior of the dark sector can leave observable signatures in the cosmic microwave background. Another important consideration is that the gravitational response of an underdense environment need not be determined by its matter content alone if the component responsible for cosmic acceleration also possesses spatial structure. Comparatively less emphasis has still been placed on the more direct and localized question of whether there could exist sufficiently extended regions of the cosmic web in which the dark energy density itself is appreciably smaller than its cosmological mean. We believe that this is a natural question deserving investigation in its own right, independently of whether such regions eventually prove to exist. \\ 

How does that, however, lead us to question the global nature of DE, is what one could ask. So, before we go any further, we would like to clarify what we are asking and what we are not asking in this work. We are not suggesting that the success of the cosmological principle should be discarded, nor are we assuming that the Universe must possess a globally anisotropic expansion history or that dark energy necessarily varies across space. Neither are we proposing that the existence of ordinary matter voids automatically implies the existence of corresponding dark energy deficiencies, since the cosmological constant remains perfectly homogeneous even in highly underdense environments. Rather, our central question is whether dark energy, if it represents a dynamical physical component rather than a strict cosmological constant, could develop sufficiently large and coherent spatial variations for certain regions to become deficient relative to the cosmological background. We are particularly interested in whether these regions could be identified through their expansion histories, peculiar velocity fields and associated gravitational observables, after accounting for the effects of ordinary matter inhomogeneities. The purpose is not to replace the successful homogeneous cosmological description, but to formulate a falsifiable test of one of its central assumptions concerning the spatial behavior of cosmic acceleration. \\

One can also think of this question in terms of historical analogies in astronomy and cosmology. Interestingly, the discovery of cosmic acceleration itself provides an instructive example of how an unexpected feature of astronomical distance measurements can reveal an otherwise unanticipated component of the cosmic energy budget. Likewise, the identification of cosmic voids demonstrated that the large scale distribution of matter is far from uniform locally, even when the Universe admits a statistically homogeneous description on sufficiently large scales \cite{deh3}. The separation of baryonic and dark matter distributions in colliding galaxy clusters \cite{deh4}, together with the identification of galaxies apparently possessing unusually little dark matter as discussed before, subsequently illustrated how exceptional astrophysical environments can reveal properties of cosmic components that remain hidden in more typical systems. Even the longstanding missing baryon problem demonstrated the importance of asking where a cosmologically established component resides, rather than assuming that its abundance must be equally apparent in every environment \cite{deh6}. These examples are not intended to suggest that dark energy must behave in the same manner as matter, since its pressure, dynamics and possible microscopic degrees of freedom are fundamentally different. They does establish the notion that the global abundance of a cosmic component does not, by itself, settle every question concerning its possible spatial distribution, unless its underlying physics specifically enforces homogeneity. Asking whether dark energy could be locally deficient is consequently a natural extension of a longstanding astronomical practice of investigating unusual environments to learn about otherwise poorly understood cosmic constituents. With the motivations established in a more firm sense, we shall now proceed to proposing some tests for the same.

\section{Dark Energy Deficiency Tests}
Having motivated the possibility of spatially deficient dark energy regions, we now turn towards identifying the observational signatures through which such regions could be differentiated from conventional matter underdensities. The central challenge here is to separate the effects of ordinary void dynamics from those arising due to genuine spatial variations in dark energy. To do this, we develop a framework connecting nonlinear void expansion, coherent peculiar velocities and the evolution of density and momentum perturbations, allowing for complementary tests of the spatial homogeneity of cosmic acceleration. We subsequently estimate the characteristic magnitudes of these effects and discuss their potential accessibility to present and forthcoming cosmological surveys. 

\subsection{Nonlinear Void Expansion and a Spatial Dark Energy Test}
We consider first the global spatially flat cosmological background and at sufficiently late times its expansion rate may be written as
\begin{equation}
H^{2}(a)=H_{0}^{2}\left[\Omega_{{\rm DE}0}F_{\rm DE}(a)+\Omega_{m0}\left(\frac{a_{0}}{a}\right)^{3}\right]
\end{equation}
Here $F_{\rm DE}(a)$ describes the homogeneous evolution of the dark energy density and satisfies $F_{\rm DE}=1$ for a cosmological constant. We will use the cosmological constant case when constructing the simplest null model, while the formalism itself may be extended to a dynamical dark energy background. A deep void does not satisfy the assumption that the total matter contrast is much smaller than unity. Instead, following the separate universe description of the central region of a void, we introduce a local scale factor $b(t)$ and a local expansion rate
\begin{equation}
\mathcal{H}_{v}(t)=\frac{\dot{b}(t)}{b(t)}
\end{equation}
The state of the local patch is characterized not only by its matter density but also by its coherent expansion relative to the cosmological background. Defining the smoothed matter contrast of the void through
\begin{equation}
1+\delta_m^{v}=\frac{\rho_m^{v}}{\bar{\rho}_m}
\end{equation}
and using $\rho_m^{v}\propto b^{-3}$ together with $\bar{\rho}_m\propto a^{-3}$ gives us the exact mass conservation relation
\begin{equation}
\mathcal{H}_{v}-H=-\frac{\dot{\delta}_m^{v}}{3\left(1+\delta_m^{v}\right)}
\label{densityvelocityexact}
\end{equation}
The coherent radial peculiar velocity of a shell of physical radius $R$ is therefore
\begin{equation}
v_{v}^{\rm coh}(R)=\left(\mathcal{H}_{v}-H\right)R
\label{coherentvelocity}
\end{equation}
and its velocity divergence satisfies
\begin{equation}
\theta_{v}^{\rm coh}=3\left(\mathcal{H}_{v}-H\right)
\end{equation}
In the linear growing mode, where $\dot{\delta}_m^{v}=Hf\delta_m^{v}$, \eqref{densityvelocityexact} reduces to
\begin{equation}
\mathcal{H}_{v}-H\simeq-\frac{1}{3}Hf\delta_m^{v}
\end{equation}
Thus, the matter density contrast and the coherent peculiar expansion are not independent descriptions of the void. For an underdensity with $\delta_m^{v}<0$, the growing mode naturally gives $\mathcal{H}_{v}>H$ and an outward coherent peculiar velocity. The central region may then be regarded as an approximately open Friedmann patch. For a void generated by a sufficiently smooth primordial matter underdensity, its local expansion in the homogeneous dark energy limit may be written in the form \cite{baushev2024empty}
\begin{equation}
\mathcal{H}_{v{\rm hom}}(b)=H_{0}\sqrt{\Omega_{{\rm DE}0}+3\varpi\Omega_{m0}\left(\frac{a_{0}}{b}\right)^{2}+\Omega_{m0}\left(\frac{a_{0}}{b}\right)^{3}}
\label{voidhom}
\end{equation}
The parameter $\varpi$ here characterizes the primordial amplitude of the matter underdensity and the second term inside the square root represents the effective open curvature generated by the initial underdensity while the final term represents the diluted matter density inside the expanding void region. We note that this description remains applicable when the final void contrast is nonlinear, which is the principal advantage over a direct linear perturbation treatment of the entire void. The parameter $\varpi$ may be defined at an early epoch $z_{1}$ during matter domination through
\begin{equation}
\varpi=(1+z_{1})\vartheta_{1}
\end{equation}
where $\vartheta_{1}>0$ denotes the magnitude of the primordial matter underdensity. The initial perturbation may be small even though its descendant at the present epoch is highly nonlinear. At the same early epoch, the growing mode fixes the corresponding peculiar expansion rather than allowing it to be chosen independently, writing the initial matter contrast as $\delta_{m1}^{v}\simeq-\vartheta_{1}$ gives
\begin{equation}
\dot{\delta}_{m1}^{v}=H_{1}f_{1}\delta_{m1}^{v}
\end{equation}
and therefore
\begin{equation}
\mathcal{H}_{v1}=H_{1}-\frac{H_{1}f_{1}\delta_{m1}^{v}}{3\left(1+\delta_{m1}^{v}\right)}
\label{initialvoidexpansion}
\end{equation}
For $|\delta_{m1}^{v}|\ll1$ this becomes
\begin{equation}
\mathcal{H}_{v1}\simeq H_{1}\left(1+\frac{f_{1}\varpi}{3(1+z_{1})}\right)
\end{equation}
so the primordial matter underdensity is accompanied by a small coherent excess expansion. The initial density and velocity perturbations therefore specify a single growing mode, with the corresponding local curvature fixed by the local Friedmann equation rather than introduced as an independent degree of freedom. The present expansion of the central void region may be related to its present matter density through the magnification factor $k=\frac{b_{0}}{a_{0}}$ and then matter conservation then gives the central matter contrast
\begin{equation}
\delta_{c}=\frac{1}{k^{3}}-1
\label{deltac}
\end{equation}
Thus, the condition $k>1$ corresponds to a region that has expanded more strongly than the cosmological background and consequently contains a lower matter density. The nonlinear matter deficiency is here encoded directly in the relative expansion of the local scale factor.
\\
\\
The value of $k$ can be obtained by demanding that the local void region and the cosmological background have the same age. For homogeneous dark energy in the cosmological constant limit this gives \cite{baushev2024empty}
\begin{equation}
\frac{\operatorname{arcosh}\left(\Omega_{m0}^{-1/2}\right)}{\sqrt{1-\Omega_{m0}}}=\frac{3}{2}\int_{0}^{k}\frac{\sqrt{x}\,dx}{\sqrt{x^{3}+\Omega_{m0}\left(1+3\varpi x-x^{3}\right)}}
\label{agematching}
\end{equation}
This relation provides us the nonlinear connection between the primordial amplitude and the present matter contrast. In particular, it does not require the final value of $\delta_c$ to be perturbatively small. The measured matter profile of a void can hence be associated with a nonlinear local expansion history before any spatial variation of dark energy is introduced. We will now generalize the local Friedmann description by allowing the dark energy density within the void region to differ from its cosmic value, and so we write
\begin{equation} \label{rhodev}
\rho_{\rm DE}^{v}(r,t)=\bar{\rho}_{\rm DE}(t)\left[1-A_{\rm DE}(t)W_{v}(r)\right]
\end{equation}
The quantity $A_{\rm DE}>0$ parametrizes a dark energy deficiency and $W_{v}$ describes its spatial profile. We normalize the central region so that $W_{v}\simeq1$ in the deepest part of the void and $W_{v}\rightarrow0$ outside the region affected by the putative deficiency. The local Friedmann equation can then be written schematically as
\begin{equation}
\mathcal{H}_{v}^{2}=\frac{8\pi G}{3}\left[\rho_{m}^{v}+\rho_{\rm DE}^{v}\right]+\frac{\mathcal{K}_{v}}{b^{2}}
\label{voidfriedmann}
\end{equation}
Here $\mathcal{K}_{v}>0$ denotes the effective open curvature contribution associated with the nonlinear matter underdensity. The corresponding local density parameters contain both the change in density and the change in the local expansion rate. In particular,
\begin{equation}
\Omega_m^{v}=\Omega_m\left(1+\delta_m^{v}\right)\left(\frac{H}{\mathcal{H}_{v}}\right)^{2}
\end{equation}
and
\begin{equation}
\Omega_{\rm DE}^{v}=\Omega_{\rm DE}\left(1-A_{\rm DE}W_v\right)\left(\frac{H}{\mathcal{H}_{v}}\right)^{2}
\end{equation}
while the effective curvature contribution may be written as
\begin{equation}
\Omega_{\mathcal{K}}^{v}=\frac{\mathcal{K}_{v}}{b^{2}\mathcal{H}_{v}^{2}}
\end{equation}
so that
\begin{equation}
\Omega_m^{v}+\Omega_{\rm DE}^{v}+\Omega_{\mathcal{K}}^{v}=1
\end{equation}
The effective cosmological parameters of the void therefore cannot be inferred from its density contrast alone, since the coherent peculiar expansion enters explicitly through $\mathcal{H}_{v}$. In the homogeneous dark energy null model the same measured matter distribution and curvature give
\begin{equation}
\mathcal{H}_{v{\rm hom}}^{2}=\frac{8\pi G}{3}\left[\rho_{m}^{v}+\bar{\rho}_{\rm DE}\right]+\frac{\mathcal{K}_{v}}{b^{2}}
\end{equation}
The comparison here must be be made between two voids with the same measured matter profile and curvature contribution. The purpose of this construction is to avoid identifying the ordinary enhanced expansion produced by the matter underdensity itself with evidence for spatially varying dark energy. Subtracting the two local Friedmann equations gives us,
\begin{equation}
\mathcal{H}_{v}^{2}-\mathcal{H}_{v{\rm hom}}^{2}=-\frac{8\pi G}{3}\bar{\rho}_{\rm DE}A_{\rm DE}W_{v}
\end{equation}
Using the global relation between the dark energy density and its density parameter gives
\begin{equation}
\mathcal{H}_{v}^{2}-\mathcal{H}_{v{\rm hom}}^{2}=-H^{2}\Omega_{\rm DE}A_{\rm DE}W_{v}
\label{Hresidualexact}
\end{equation}
This is the central expansion test for us and we shall note here why. A dark energy deficient region has a smaller local expansion rate than a void with the same matter content evolving in a homogeneous dark energy background. The observable is therefore not whether a void expands more rapidly than the cosmic average, but whether it expands less rapidly than expected from its nonlinear matter deficiency. For a sufficiently small deficiency it is easy to check that,
\begin{equation}
\frac{\Delta\mathcal{H}_{v}}{\mathcal{H}_{v{\rm hom}}}\simeq-\frac{1}{2}\frac{H^{2}\Omega_{\rm DE}}{\mathcal{H}_{v{\rm hom}}^{2}}A_{\rm DE}W_{v}
\label{Hresiduallinear}
\end{equation}
The response is reduced relative to $A_{\rm DE}/2$ because the expansion of a void is already partially supported by its open curvature contribution. This makes the matter and curvature reconstruction essential for translating a measured local Hubble residual into a dark energy deficiency. We also note that equation \eqref{Hresiduallinear} is most naturally interpreted as an observational comparison at fixed measured void profile. In a complete physical model the spatially varying dark energy also modifies the past evolution of the matter distribution and hence the present value of $b_{0}$. The self consistent solution should therefore satisfy the age matching condition
\begin{equation}
\int_{0}^{a_{0}}\frac{da}{aH(a)}=\int_{0}^{b_{0}}\frac{db}{b\mathcal{H}_{v}(b)}
\end{equation}
A specified model for $\rho_{\rm DE}^{v}(b)$ can hence be evolved from the primordial void perturbation to the observed epoch. The simpler fixed profile comparison provides a model independent phenomenological test, while this equation supplies the route toward a fully dynamical interpretation and the observed line of sight motion also contains contributions unrelated to the local void expansion. After transformation to an appropriate cosmological rest frame, it is useful to write this as
\begin{equation}
v_{\rm los}^{\rm obs}=\mathbf{V}_{\rm bulk}\cdot\hat{\mathbf{n}}+\left(\mathcal{H}_{v}-H\right)R\mu_{v}+u_{\parallel}
\label{velocitydecomposition}
\end{equation}
Here $\mathbf{V}_{\rm bulk}$ is the coherent motion of the larger environment, $R$ is the physical distance from the void center, $\mu_{v}$ describes its projection along the line of sight and $u_{\parallel}$ contains residual velocity fluctuations. The second contribution is the local radial motion generated by the different expansion history of the void. The coherent term in Eq. \eqref{velocitydecomposition} is not independent of the smoothed matter contrast, but is related to it through Eq. \eqref{densityvelocityexact}. The density profile therefore determines the mass conservation expectation for the coherent flow, while peculiar velocity observations provide an independent measurement of the actual local expansion rate. This separation is important observationally, since a dark energy deficiency is identified through a mismatch between the measured coherent expansion and the homogeneous dark energy prediction for the same reconstructed matter profile. Under the homogeneous dark energy null hypothesis the corresponding coherent velocity is
\begin{equation}
v_{v{\rm hom}}^{\rm los}=\left(\mathcal{H}_{v{\rm hom}}-H\right)R\mu_{v}
\end{equation}
The coherent residual produced by a dark energy deficiency is 
\begin{equation}
\Delta v_{v}^{\rm los}=\left(\mathcal{H}_{v}-\mathcal{H}_{v{\rm hom}}\right)R\mu_{v}
\label{velocityresidual}
\end{equation}
For an approximately spherical region, the measured coherent peculiar velocity divergence may equivalently be used to write
\begin{equation}
\mathcal{H}_{v}=H+\frac{\theta_{v}^{\rm coh}}{3}
\end{equation}
and the effective deficiency inferred from the expansion residual is then
\begin{equation}
A_{\rm DE}W_v=\frac{\mathcal{H}_{v{\rm hom}}^{2}-\left(H+\theta_{v}^{\rm coh}/3\right)^{2}}{H^{2}\Omega_{\rm DE}}
\label{velocitydeficiencyestimator}
\end{equation}
This relation makes the proposed null test directly operational, as the matter profile and growing mode determine $\mathcal{H}_{v{\rm hom}}$, while the coherent peculiar velocity field determines the actual $\mathcal{H}_{v}$. This means that any statistically significant residual is then interpreted only after the ordinary matter induced expansion has been accounted for. A stacked void analysis can search for this radial pattern after marginalizing over the bulk flow and the ordinary nonlinear void motion. The sign and radial dependence are more informative than a single peculiar velocity measurement because a genuine void centered signal must reverse projection across the void center and follow the void geometry. The scalar expansion description is most appropriate for isolated and approximately spherical voids. A more general region may instead be described by the symmetric velocity gradient tensor given by
\begin{equation}
\Theta_{ij}^{v}=\frac{1}{2}\left(\frac{\partial v_{i}}{\partial R_{j}}+\frac{\partial v_{j}}{\partial R_{i}}\right)
\end{equation}
Its trace defines the local volume expansion given as 
\begin{equation}
\mathcal{H}_{\rm loc}=\frac{1}{3}{\rm Tr}\Theta^{v}
\end{equation}
while its traceless component defines the shear
\begin{equation}
\sigma_{ij}^{v}=\Theta_{ij}^{v}-\mathcal{H}_{\rm loc}\delta_{ij}
\end{equation}
Interacting or colliding voids may possess significant shear and can produce walls, pancakes and filamentary flows that cannot be captured by a single $\mathcal{H}_{v}$. A first observational implementation should hence preferentially select isolated voids with small ellipticity and weak environmental shear, while the tensor description provides a natural extension to nonspherical systems.

\subsection{Residual Density and Momentum Power Spectra}
The nonlinear nature of the background void does not prevent a perturbative treatment of smaller residual fluctuations about that background. We hence decompose the densities locally as $\rho_{m}=\rho_{m}^{v}\left(1+\widehat{\delta}_{m}\right)$ and $\rho_{\rm DE}=\rho_{\rm DE}^{v}\left(1+\widehat{\Delta}_{\rm DE}\right)$.
The quantities $\widehat{\delta}_{m}$ and $\widehat{\Delta}_{\rm DE}$ describe fluctuations about the already nonlinear smoothed void background. They may remain perturbatively small even when the matter contrast of the complete void relative to the cosmic mean is close to $-1$. This distinction allows linear perturbation methods to be used without requiring the void itself to be a linear structure. At this stage it is important to distinguish the coherent void velocity already contained in the background evolution from the smaller residual velocity field. We therefore write schematically
\begin{equation}
\mathbf{v}_{\rm pec}=\mathbf{v}_{v}^{\rm coh}+\mathbf{u}
\end{equation}
where $\mathbf{v}_{v}^{\rm coh}$ satisfies the nonlinear mass conservation relation \eqref{densityvelocityexact}, while $\mathbf{u}$ describes perturbations about that coherent flow. The continuity and Euler equations below consequently describe the residual perturbations around a background whose density and peculiar expansion are already dynamically consistent. Let $\mathbf{u}$ denote the residual velocity after the coherent nonlinear void flow and bulk motion have been removed and we define its divergence through $\theta_{v}=\nabla\cdot\mathbf{u}$. The linear continuity equation in the local void background is then given as
\begin{equation}
\dot{\widehat{\delta}}_{m}+\frac{\theta_{v}}{b}=0
\end{equation}
The corresponding Euler equation is
\begin{equation}
\dot{\theta}_{v}+\mathcal{H}_{v}\theta_{v}=-\frac{1}{b}\nabla^{2}\phi_{v}
\end{equation}
The local gravitational potential receives contributions from both matter and dark energy fluctuations. In the regime where a local Poisson description is appropriate we may write
\begin{equation}
\nabla^{2}\phi_{v}=4\pi Gb^{2}\left[\rho_{m}^{v}\widehat{\delta}_{m}+\rho_{\rm DE}^{v}\widehat{\Delta}_{\rm DE}\right]
\end{equation}
This equation provides the second manner in which spatial dark energy structure affects peculiar velocities. The mean deficiency changes the nonlinear background expansion $\mathcal{H}_{v}$, while fluctuations about that mean contribute directly to the residual gravitational field. Combining the local continuity and Euler equations gives us,
\begin{equation}
\ddot{\widehat{\delta}}_{m}+2\mathcal{H}_{v}\dot{\widehat{\delta}}_{m}-4\pi G\rho_{m}^{v}\widehat{\delta}_{m}=4\pi G\rho_{\rm DE}^{v}\widehat{\Delta}_{\rm DE}
\label{coupledgrowth}
\end{equation}
This result is particularly important for the interpretation of the velocity power spectrum. Dark energy fluctuations do not simply add a second velocity contribution to an otherwise unchanged matter power spectrum. They also modify the evolution of the matter fluctuations themselves, so the measured $P_{mm}^{v}$ must be treated as the product of the coupled evolution described by \eqref{coupledgrowth}. The formal solution may be separated into the solution obtained without local dark energy fluctuations and the response sourced by them as $\widehat{\delta}_{m}=\widehat{\delta}_{m}^{(0)}+\widehat{\delta}_{m}^{({\rm DE})}$, the dark energy induced matter response satisfies
\begin{equation}
\ddot{\widehat{\delta}}_{m}^{({\rm DE})}+2\mathcal{H}_{v}\dot{\widehat{\delta}}_{m}^{({\rm DE})}-4\pi G\rho_{m}^{v}\widehat{\delta}_{m}^{({\rm DE})}=4\pi G\rho_{\rm DE}^{v}\widehat{\Delta}_{\rm DE}
\end{equation}
Introducing the corresponding retarded growth Green function gives
\begin{equation}
\widehat{\delta}_{m}^{({\rm DE})}(\mathbf{k}t)=4\pi G\int_{t_{i}}^{t}G_{v}(tt')\rho_{\rm DE}^{v}(t')\widehat{\Delta}_{\rm DE}(\mathbf{k}t')dt'
\end{equation}
The residual matter power spectrum therefore contains an integrated memory of the dark energy perturbation history. This is why the appropriate observable is a joint density and momentum statistic rather than a velocity power spectrum interpreted against a fixed $\Lambda$CDM matter template. We introduce the dimensionless local velocity divergence,
\begin{equation}
\vartheta_{v}=-\frac{\theta_{v}}{b\mathcal{H}_{v}}
\end{equation}
The most general linear two source response may then be written phenomenologically as
\begin{equation}
\vartheta_{v}=\mathcal{G}_{m}^{v}\widehat{\delta}_{m}+\mathcal{G}_{\rm DE}^{v}\widehat{\Delta}_{\rm DE}
\label{twosource}
\end{equation}
The functions $\mathcal{G}_{m}^{v}$ and $\mathcal{G}_{\rm DE}^{v}$ encode the local dynamical response of the velocity field. They may depend on scale, epoch and the nonlinear background void parameters. In the homogeneous dark energy single source limit one recovers the usual deterministic relation between the matter density and velocity divergence. The residual velocity field in Fourier space is,
\begin{equation}
\mathbf{u}(\mathbf{k})=i\frac{b\mathcal{H}_{v}\mathbf{k}}{k^{2}}\vartheta_{v}(\mathbf{k})
\end{equation}
Its line of sight component is therefore
\begin{equation}
u_{\parallel}(\mathbf{k})=i\frac{b\mathcal{H}_{v}\mu}{k}\vartheta_{v}(\mathbf{k})
\end{equation}
The factor $1/k$ makes peculiar velocity statistics especially sensitive to large spatial scales. This is useful for a search for spatial dark energy structure because any physically viable dark energy inhomogeneity is expected to become most relevant on comparatively large scales. The local momentum field can be defined by
\begin{equation}
p_{v}=\left(1+\widehat{\delta}_{g}\right)u_{\parallel}
\end{equation}
At linear order this reduces to $p_{v}\simeq u_{\parallel}$, and the environment conditioned momentum power spectrum satisfies
\begin{equation}
P_{pp}^{v}(k,\mu)=\frac{\left(b\mathcal{H}_{v}\right)^{2}\mu^{2}}{k^{2}}P_{\vartheta\vartheta}^{v}(k)
\label{momentumpower}
\end{equation}
This has the same characteristic expansion and velocity structure exploited in present peculiar velocity power spectrum measurements, but the expansion rate and perturbation transfer functions are now those of the nonlinear void environment rather than those of the global homogeneous background. Using \eqref{twosource} gives us,
\begin{equation}
P_{\vartheta\vartheta}^{v}=\left(\mathcal{G}_{m}^{v}\right)^{2}P_{mm}^{v}+\left(\mathcal{G}_{\rm DE}^{v}\right)^{2}P_{{\rm DE}{\rm DE}}^{v}+2\mathcal{G}_{m}^{v}\mathcal{G}_{\rm DE}^{v}P_{m{\rm DE}}^{v}
\label{velocitypowersources}
\end{equation}
The first term contains the fully evolved matter power spectrum inside the void environment. The second is the direct contribution from dark energy fluctuations and the third represents their correlation. All three terms are physically relevant because the matter and dark energy sectors need not fluctuate independently. The matter velocity cross spectrum follows as
\begin{equation}
P_{m\vartheta}^{v}=\mathcal{G}_{m}^{v}P_{mm}^{v}+\mathcal{G}_{\rm DE}^{v}P_{m{\rm DE}}^{v}
\end{equation}
A particularly useful environment dependent effective response can therefore be constructed directly from observables
\begin{equation}
f_{\rm eff}^{v}(k)=\frac{P_{m\vartheta}^{v}(k)}{P_{mm}^{v}(k)}
\label{feffvoid}
\end{equation}
This quantity tests whether the observed residual velocity field follows the relation expected from the measured matter distribution. A dark energy fluctuation correlated with matter can modify $f_{\rm eff}^{v}$ even if it does not generate an independent stochastic velocity component. To isolate any contribution that cannot be deterministically reconstructed from the matter field, we define
\begin{equation}
P_{\rm extra}^{v}=P_{\vartheta\vartheta}^{v}-\frac{\left(P_{m\vartheta}^{v}\right)^{2}}{P_{mm}^{v}}
\label{extravelocity}
\end{equation}
For a perfectly deterministic single source relation this quantity vanishes in the ideal linear limit. A positive residual can indicate an additional source of velocity fluctuations, although nonlinear structure etc. must also be modeled before such an excess can be attributed to dark energy. The same information may be expressed through the density velocity coherence coefficient
\begin{equation}
r_{m\vartheta}^{v}(k)=\frac{P_{m\vartheta}^{v}(k)}{\sqrt{P_{mm}^{v}(k)P_{\vartheta\vartheta}^{v}(k)}}
\end{equation}
The ideal deterministic limit gives $r_{m\vartheta}^{v}=1$. An additional component uncorrelated with matter produces $r_{m\vartheta}^{v}<1$. A measured coherence consistent with unity would not, however, exclude dark energy inhomogeneity because a dark energy component strongly correlated with matter would instead appear through the scale and environment dependence of $f_{\rm eff}^{v}$. The observational construction should therefore compare void regions with matched control environments. Schematically one may consider writng that,
\begin{equation}
\Delta f_{\rm env}(k)=f_{\rm eff}^{v}(k)-f_{\rm eff}^{c}(k)
\end{equation}
where the control sample is selected to minimize differences unrelated to the large scale environment. A spatial dark energy effect should appear coherently in the void conditioned expansion, momentum power and density velocity relation rather than solely in one individual statistic. The density, momentum and density momentum spectra provide a natural data vector
\begin{equation}
\mathbf{D}_{v}=\left\{P_{0}^{\delta\delta v},P_{2}^{\delta\delta v},P_{0}^{ppv},P_{1}^{\delta pv}\right\}
\end{equation}
These are the same classes of observables that can already be measured from modern density and peculiar velocity surveys \cite{qin2026desi}. In practice the environment selection produces an additional window function and the theoretical spectra must therefore be convolved with both the survey geometry and the void selection function before comparison with the data. The two manifestations of the proposed signal are consequently connected and a mean dark energy deficiency changes $\mathcal{H}_{v}$ and produces a coherent radial velocity residual through \eqref{velocityresidual}, while dark energy fluctuations and their effect on matter growth modify the residual momentum spectra through \eqref{velocitypowersources}. The two measurements should therefore be treated jointly rather than as unrelated searches.

\subsection{Illustrative Magnitudes and Survey Sensitivity}

We now consider simple numerical examples in order to estimate the characteristic size of these effects. We adopt representative cosmological values close to those used in the DESI DR1 peculiar velocity analysis \cite{qin2026desi}
\begin{equation}
H_{0}=67.36\ {\rm km\ s^{-1}\ Mpc^{-1}}, \quad \Omega_{m0}=0.301 , \quad \Omega_{{\rm DE}0}=0.699
\end{equation}
Consider a representative nonlinear central void characterized by $\varpi=1.45$, solving the age matching relation in Eq. \eqref{agematching} gives approximately
\begin{equation}
k\simeq1.55
\end{equation}
The corresponding central matter contrast is
\begin{equation}
\delta_{c}\simeq\frac{1}{1.55^{3}}-1\simeq-0.73
\end{equation}
This example hence corresponds to a genuinely nonlinear void and lies well outside the regime in which the total void may consistently be regarded as a small density perturbation. Nevertheless, its central expansion can be described through the separate universe construction. Using \eqref{voidhom} gives,
\begin{equation}
\frac{\mathcal{H}_{v{\rm hom}}}{H_{0}}\simeq\sqrt{0.699+\frac{3(1.45)(0.301)}{1.55^{2}}+\frac{0.301}{1.55^{3}}}\simeq1.152
\end{equation}
The corresponding local expansion rate is
\begin{equation}
\mathcal{H}_{v{\rm hom}}\simeq77.6\ {\rm km\ s^{-1}\ Mpc^{-1}}
\end{equation}
The same example also illustrates why the density and peculiar expansion information should be retained simultaneously. The coherent excess expansion of the homogeneous dark energy void relative to the global background is
\begin{equation}
\mathcal{H}_{v{\rm hom}}-H_{0}\simeq10.24\ {\rm km\ s^{-1}\ Mpc^{-1}}
\end{equation}
which corresponds to coherent outward velocities of approximately
\begin{equation}
v_{v{\rm hom}}^{\rm coh}(30\ {\rm Mpc})\simeq307\ {\rm km\ s^{-1}}
\end{equation}
and
\begin{equation}
v_{v{\rm hom}}^{\rm coh}(50\ {\rm Mpc})\simeq512\ {\rm km\ s^{-1}}
\end{equation}
before projection effects are included. The corresponding effective local density parameters in the homogeneous dark energy null model are approximately
\begin{equation}
\Omega_{m}^{v}\simeq0.061 , \quad \Omega_{\rm DE}^{v}\simeq0.527 , \quad \Omega_{\mathcal{K}}^{v}\simeq0.412
\end{equation}
These values show explicitly that the effective local cosmology reflects both the reduced matter density and the enhanced coherent expansion of the void. The enhanced expansion relative to $H_{0}$ is generated entirely by the nonlinear matter deficiency and its associated curvature. It therefore represents the null expectation rather than evidence for a dark energy effect. Now suppose that the central dark energy density is deficient by ten percent relative to the cosmic mean
\begin{equation}
A_{\rm DE}=0.10
\end{equation}
At fixed observed matter profile and curvature, Eq. \eqref{Hresidualexact} gives
\begin{equation}
\frac{\mathcal{H}_{v}^{2}}{H_{0}^{2}}\simeq1.152^{2}-0.699(0.10)
\end{equation}
This gives us,
\begin{equation}
\mathcal{H}_{v}\simeq75.5\ {\rm km\ s^{-1}\ Mpc^{-1}}
\end{equation}
The expansion residual is therefore
\begin{equation}
\Delta\mathcal{H}_{v}\simeq-2.1\ {\rm km\ s^{-1}\ Mpc^{-1}}
\end{equation}
A ten percent dark energy deficiency consequently produces an approximately $2.7\%$ suppression relative to the homogeneous dark energy void expansion in this example. The effect is considerably smaller than the total enhancement of the void expansion itself but can become measurable statistically if the null void dynamics are reconstructed sufficiently accurately. The distinction between the total coherent void velocity and the dark energy induced residual is therefore important. For the $30\ {\rm Mpc}$ example, the homogeneous dark energy matter void produces an outward coherent flow of order $307\ {\rm km\ s^{-1}}$, while a ten percent dark energy deficiency changes this prediction by only about $62\ {\rm km\ s^{-1}}$. The observational test is consequently a residual test around the matter and velocity consistent nonlinear null solution rather than a search for an outward peculiar velocity by itself. 
\\
\\
For a five percent deficiency one similarly obtains
\begin{equation}
A_{\rm DE}=0.05 , \quad \Delta\mathcal{H}_{v}\simeq-1.0\ {\rm km\ s^{-1}\ Mpc^{-1}}
\end{equation}
The scaling is approximately linear for modest deficiency amplitudes and this makes stacked measurements of similar voids particularly useful because the coherent expansion residual can accumulate statistically while random peculiar motions average down. For a characteristic physical radius of $30\ {\rm Mpc}$ the ten percent example produces a radial velocity shift of order
\begin{equation}
\left|\Delta v_{v}\right|\simeq\left|\Delta\mathcal{H}_{v}\right|R\simeq62\ {\rm km\ s^{-1}}
\end{equation}
For a larger region with radius $50\ {\rm Mpc}$ the corresponding estimate becomes
\begin{equation}
\left|\Delta v_{v}\right|\simeq104\ {\rm km\ s^{-1}}
\end{equation}
These numbers represent the maximal radial signal before projection and profile averaging and the observed line of sight amplitude will be smaller for individual galaxies, but the distinctive radial pattern can be recovered statistically by stacking large samples of suitably isolated voids. The local expansion also appears explicitly in the prefactor of the momentum power spectrum and if the dimensionless response $P_{\vartheta\vartheta}^{v}$ is temporarily held fixed, Eq. \eqref{momentumpower} gives
\begin{equation}
\frac{P_{pp}^{v}}{P_{pp{\rm hom}}^{v}}\simeq\left(\frac{\mathcal{H}_{v}}{\mathcal{H}_{v{\rm hom}}}\right)^{2}
\end{equation}
For the ten percent deficiency considered above this becomes,
\begin{equation}
\frac{P_{pp}^{v}}{P_{pp{\rm hom}}^{v}}\simeq0.947
\end{equation}
The altered local expansion alone therefore produces approximately a five percent change in the momentum power normalization. The complete signal can differ from this estimate because $P_{\vartheta\vartheta}^{v}$ itself changes through the modified matter growth, direct dark energy fluctuations and the matter dark energy cross spectrum. The DESI DR1 power spectrum analysis currently obtains \cite{qin2026desi}
\begin{equation}
f\sigma_{8}(z_{\rm eff}=0.07)=0.440^{+0.080}_{-0.096}
\end{equation}
while the consensus combination of the DESI DR1 peculiar velocity analyses gives us,
\begin{equation}
f\sigma_{8}(z_{\rm eff}=0.07)=0.450\pm0.055
\end{equation}
The present global precision is therefore at approximately the ten percent level for the consensus growth measurement, while an environment conditioned void analysis will initially be less precise because only a subset of the survey volume and peculiar velocity sample can be used. The planned DESI peculiar velocity sample has been forecast to contain approximately $1.3\times10^{5}$ fundamental plane velocities and approximately $5\times10^{4}$ Tully Fisher velocities, with a forecast global statistical precision of order four percent on the growth rate at low redshift \cite{saulder2023target}. This does not directly imply four percent precision for a void conditioned measurement, but it establishes that substantially more precise density velocity statistics are expected to become available. If a future void conditioned analysis achieved a fractional uncertainty $\epsilon_{\mathcal{H}}$ on the stacked local expansion, Eq. \eqref{Hresiduallinear} gives  us approximately,
\begin{equation}
A_{\rm DE}^{5\sigma}\simeq\frac{10\epsilon_{\mathcal{H}}\mathcal{H}_{v{\rm hom}}^{2}}{H^{2}\Omega_{\rm DE}}
\end{equation}
For the representative void considered above the response coefficient is
\begin{equation}
\frac{1}{2}\frac{H_{0}^{2}\Omega_{{\rm DE}0}}{\mathcal{H}_{v{\rm hom}}^{2}}\simeq0.263
\end{equation}
A one percent measurement of the stacked local expansion would consequently require approximately $A_{\rm DE}^{5\sigma}\simeq0.19$ and a half percent measurement would instead give
$A_{\rm DE}^{5\sigma}\simeq0.095$, while a two tenths of a percent determination would formally reach $A_{\rm DE}^{5\sigma}\simeq0.038$. These values are schematic sensitivity targets rather than survey forecasts and they show, however, that percent and subpercent measurements of the mean expansion of stacked nonlinear void environments would begin to probe dark energy deficiencies at the few to ten percent level. A complementary estimate can be made for the momentum statistics and since the linear momentum power normalization scales approximately as the square of the velocity response, a fractional uncertainty $\epsilon_{f}$ on an effective growth amplitude corresponds approximately to
\begin{equation}
\epsilon_{P_{pp}}\simeq2\epsilon_{f}
\end{equation}
A four percent determination of the relevant velocity response would therefore correspond to an approximately eight percent measurement of the momentum power normalization before the information from the density momentum cross spectrum is included. The five percent momentum normalization shift generated by the local expansion alone in the ten percent deficiency example would end up be challenging by itself, but the additional scale dependent terms in Eq. \eqref{velocitypowersources} provide further information that is absent from a pure amplitude measurement. Another important consistency requirement arises from the observed matter power spectrum itself, since spatial dark energy perturbations can modify the gravitational potentials governing the growth of matter fluctuations \cite{gpde1sapone2010fingerprinting,gpde2sapone2009fingerprinting,gpde3sapone2012fingerprinting,gpde4novell2025full}. This provides us with an independent constraint on the amplitude and spatial abundance of dark energy deficient regions, beyond the peculiar velocity signatures considered so far. To quantify the relative importance of the additional gravitational source, we may introduce the dimensionless ratio
\begin{equation}
\epsilon_{\mathrm{DE}}(k,z) = \frac{\Omega_{\mathrm{DE}}(z)}{\Omega_m(z)}
\frac{\Delta_{\mathrm{DE}}(k,z)}
{\delta_m(k,z)}
\end{equation}
where $\Delta_{\mathrm{DE}}$ denotes the relevant dark energy density perturbation in the gravitational source equation. For our representative void with $\delta_m^v\simeq -0.73$, a local dark energy deficiency of amplitude $A_{\mathrm{DE}}=0.1$-$0.2$ would correspond to a source contribution approximately $0.32$-$0.64$ times that of the matter perturbation at the present epoch, within the simplified Poisson description. Such a contribution would be potentially significant for local matter dynamics, although it should not be identified directly with the fractional change in the matter power spectrum as that depends on the integrated growth history, spatial correlations and the distribution of deficient regions. Since the observed galaxy clustering and peculiar-velocity measurements over $0<z<1$ remain broadly consistent with $\Lambda$CDM, any proposed population of dark energy deficient regions must reproduce the corresponding matter clustering statistics within their observational uncertainties. More explicitly, the additional matter response generated by dark energy perturbations modifies the power spectrum according to
\begin{equation}
\Delta P_{mm}^{\mathrm{DE}} = 2P_{m^{(0)}m^{(\mathrm{DE})}} + P_{m^{(\mathrm{DE})}m^{(\mathrm{DE})}}
\end{equation}
where $\delta_m^{(0)}$ denotes the matter perturbation in the homogeneous dark energy reference model and $\delta_m^{(\mathrm{DE})}$ represents the additional response sourced by dark energy fluctuations. Consistency with observations consequently would require $|\Delta P_{mm}^{\mathrm{DE}}/P_{mm}^{\Lambda\mathrm{CDM}}|$ to remain below the scale and redshift dependent observational tolerance, after accounting for the covariance and uncertainties in the underlying cosmological parameters. As a simple illustration, assuming a coherent relation $\Delta_{\mathrm{DE}}=Q\delta_m$ with $Q=A_{\mathrm{DE}}/0.73$ operating from $z=1$ to the present, a linear growth calculation gives an approximately $5\%$ enhancement in the matter power spectrum for $A_{\mathrm{DE}}=0.2$. This estimate is not a survey derived upper bound, but demonstrates that even moderately large dark energy deficiencies need not automatically produce order unity departures from the observed clustering amplitude. The actual constraint will depend strongly on the characteristic void radius, the redshift evolution and sound speed of dark energy, the fraction of the cosmic volume occupied by deficient regions and their correlations with the matter distribution. In particular, a rare localized deficiency may remain compatible with global clustering measurements even when its local dynamical influence is substantial. A definitive upper limit on $A_{\mathrm{DE}}$ would require evolving the coupled matter and dark energy perturbations and confronting the resulting scale dependent power spectra with large scale structure observations, which we would leave for a future work.
\\
\\
Future peculiar velocity information need not arise from a single survey, as later DESI samples can enlarge the Fundamental Plane and Tully Fisher catalogues, while supernova distances provide an independent route to peculiar velocities. Forecast studies for the Rubin Observatory supernova sample indicate sensitivity to low redshift growth through supernova peculiar velocities \cite{rosselli2025forecast} and combining independent distance indicators is particularly useful for the present problem because common environmental residuals can be tested against different velocity calibration systematics. A particularly important consistency requirement is that the reconstructed matter profile and the measured coherent peculiar velocity field should first satisfy the nonlinear mass conservation relation \eqref{densityvelocityexact} within the homogeneous dark energy null model. The dark energy search then concerns the residual departure from this jointly density and velocity calibrated prediction rather than a comparison based on density alone. The strongest implementation of this framework would hence simultaneously reconstruct the nonlinear matter profile and homogeneous dark energy void solution, measure the coherent residual $\Delta\mathcal{H}_{v}$ and fit the void conditioned density, momentum and cross spectra. A dark energy deficiency would then be required to explain both the mean expansion anomaly and the scale dependent residual velocity statistics. The gravitational potential evolution of the same void population, measured through lensing and the integrated Sachs Wolfe effect, can subsequently provide an independent confirmation channel. Agreement among these observables would test the spatial uniformity of cosmic acceleration without requiring the nonlinear void itself to be treated within linear perturbation theory.

\section{Theoretical Implications}
The phenomenological profile introduced in \eqref{rhodev} requires dark energy to possess a genuine spatial degree of freedom. At the level of an effective field theory, a sufficiently general starting point is
\begin{equation}
S=\int d^{4}x\sqrt{-g}\left[\frac{M_{\rm Pl}^{2}}{2}R+P(\phi,X)\right]+S_{m}
\end{equation}
where $X=-\frac{1}{2}g^{\mu\nu}\partial_{\mu}\phi\partial_{\nu}\phi$
and the dark energy pressure and density are
\begin{equation}
p_{\rm DE}=P , \quad \rho_{\rm DE}=2XP_{,X}-P
\end{equation}
The corresponding equation of state and physical sound speed are
\begin{equation}
w_{\rm DE}=\frac{P}{2XP_{,X}-P} , \quad c_{s}^{2}=\frac{P_{,X}}{P_{,X}+2XP_{,XX}}
\end{equation}
A theory capable of producing the regions considered here hence must simultaneously generate late time acceleration, permit nonzero dark energy perturbations and allow those perturbations to remain coherent on scales comparable with the void radius. This would require $w_{\rm DE}<-1/3$ at late times, while $w_{\rm DE}$ cannot remain identically equal to $-1$ if ordinary dark energy perturbations are to be generated. The sound horizon must also be sufficiently small that pressure support does not erase the spatial variation on the relevant scale, which would end up schematically requiring
\begin{equation}
\frac{c_s}{aH}\lesssim R_v
\end{equation}
together with the usual stability requirements $P_{,X}>0$ and $c_s^2>0$. A strict cosmological constant does not satisfy these requirements, as its stress energy tensor is
\begin{equation}
T_{\mu\nu}^{\Lambda}=-\rho_{\Lambda}g_{\mu\nu}
\end{equation}
with $w=-1$ and constant $\rho_{\Lambda}$, so there is no independent propagating degree of freedom capable of producing a profile such as \eqref{rhodev}. In standard general relativity one hence has $\delta\rho_{\Lambda}=0$ and an $A_{\rm DE}\neq0$ region necessarily requires physics beyond a strict cosmological constant. This makes dynamical dark energy particularly interesting in the present context. We know that recent DESI analyses have strengthened the motivation for an evolving equation of state, although the present evidence remains suggestive rather than conclusive and some of the newest DESI measurements remain closer to the $\Lambda$CDM expectation \cite{desi2025dr2,desi2025lya}. The possibility that dark energy is dynamical still opens precisely the additional degree of freedom required for spatial dark energy structure.
\\
\\
A simple realization which could be of use here, would instead be something like clustering k essence, for which
\begin{equation}
P(\phi,X)=\frac{X^{n}}{M^{4n-4}}-V(\phi)
\end{equation}
gives us,
\begin{equation}
c_s^2=\frac{1}{2n-1}
\end{equation}
for the kinetic contribution. For sufficiently large $n$, the sound speed can be substantially smaller than unity while the potential remains capable of driving cosmic acceleration. A slowly varying potential such as
\begin{equation}
V(\phi)=V_0\exp\left(-\lambda\frac{\phi}{M_{\rm Pl}}\right)
\end{equation}
then provides a simple proof of principle, as if the dark energy perturbation follows the matter environment through a response function $Q(k,a)$,
\begin{equation}
\delta_{\rm DE}^{v}(k,a)\simeq Q(k,a)\delta_m^{v}(k,a)
\end{equation}
then a matter void with $\delta_m^{v}<0$ produces a dark energy deficiency whenever $Q>0$. In real space this can reproduce \eqref{rhodev} through
\begin{equation}
A_{\rm DE}W_v(r)\simeq-Q\delta_m^{v}(r)
\end{equation}
provided the dark energy sound horizon is sufficiently small compared with the size of the region. Canonical uncoupled quintessence, for which $c_s^2=1$, is generally much smoother on subhorizon scales and would hence have considerably greater difficulty producing an order tens of percent deficiency across the scales considered here. A second possibility would be an environmentally or matter coupled scalar field, where one may consider
\begin{equation}
S=\int d^{4}x\sqrt{-g}\left[\frac{M_{\rm Pl}^{2}}{2}R-\frac{1}{2}(\partial\phi)^2-V(\phi)\right]+S_m\left[A^{2}(\phi)g_{\mu\nu},\psi_m\right]
\end{equation}
for which the scalar evolves in an effective potential that depends on the local matter density,
\begin{equation}
V_{\rm eff}(\phi)=V(\phi)+\rho_m A(\phi)
\end{equation}
As a simple example one may take
\begin{equation}
V(\phi)=V_0\exp\left(-\lambda\frac{\phi}{M_{\rm Pl}}\right)
\end{equation}
and
\begin{equation}
A(\phi)=\exp\left(\beta\frac{\phi}{M_{\rm Pl}}\right)
\end{equation}
so that the local equilibrium value satisfies
\begin{equation}
V_{,\phi}+\rho_m A_{,\phi}=0
\end{equation}
A matter underdensity would then shift the local scalar configuration relative to the cosmic background. For an appropriate branch of the potential and coupling this can lower the local potential energy, producing
\begin{equation}
\rho_{\rm DE}^{v}\simeq V(\phi_v)<V(\bar{\phi})\simeq\bar{\rho}_{\rm DE}
\end{equation}
and hence a profile of the form given in \eqref{rhodev}. In this class of models the characteristic radius of the dark energy response need not equal the matter void radius, since it is also controlled by the effective scalar mass and coupling. A large coherent region requires the scalar Compton wavelength in the underdense environment to be comparable to or larger than the relevant spatial scale. 
\\
\\
These examples tell us that the principal microphysical requirements are a dynamical dark energy degree of freedom, sufficiently weak pressure support or an environment dependent interaction, stability against ghost and gradient instabilities and a mechanism capable of producing $\delta_{\rm DE}<0$ in matter underdensities. Obtaining the amplitudes suggested by the phenomenological analysis, particularly $A_{\rm DE}\sim0.1$-$0.2$ over scales of several hundred Mpc, would place nontrivial restrictions on the allowed potential, sound speed and coupling strength. Fifth force constraints, screening, background expansion constraints etc. would also have to be satisfied. The two constructions above should hence be regarded only as preliminary proof of principle examples showing how \eqref{rhodev} could arise. A systematic classification of dark energy theories capable of producing stable dark energy deficient regions, together with their detailed perturbation dynamics and observational constraints, is left for future work.
\\
\\
An additional theoretical consideration concerns the nature and dynamical stability of the boundary separating a putative dark energy deficient region from its surrounding cosmological environment. One may ask whether a sufficiently sharp spatial transition in the dark energy density necessarily implies the formation of a domain wall. We emphasize that this need not be the case in the phenomenological framework considered here, as conventional domain walls arise when a scalar field interpolates between distinct vacuum configurations, typically associated with multiple minima of an underlying potential. On the other hand, a spatial variation in dark energy density does not by itself establish the existence of such disconnected vacua \cite{dw1kobayashi2026domain,dw2gruber2024domain,dw3avelino2020parameter,dw4friedland2003domain}. In clustering $k$-essence or environmentally coupled scalar field models, the local field configuration may respond continuously to the matter distribution, producing a smooth and finite width density contrast without requiring a topological defect. If the scalar evolves along a single dynamical branch or follows a unique density dependent minimum of its effective potential, the transition may be supported by ordinary field gradients rather than a localized domain wall solution. The profile $W_v(r)$ introduced in our phenomenological construction should hence be interpreted as an illustrative spatial response function, rather than an assumption that two distinct vacuum states are separated by a physical wall. \\

On the other hand, if a dark energy deficient region corresponds to a genuine vacuum bubble, with its interior and exterior occupying distinct minima of a scalar potential, the associated domain wall would introduce additional physical requirements. A conventional wall possesses surface tension and a localized contribution to the energy-momentum tensor, while its motion is governed by the competition between surface tension, the vacuum energy difference, spacetime curvature and cosmological damping. Such walls need not propagate at the speed of light, although sufficiently large vacuum pressure differences can accelerate them to relativistic velocities, potentially changing the spatial extent of the deficient region over cosmological timescales. In this situation, the assumption of a fixed comoving deficiency profile would no longer be generally justified and the evolution of the boundary would have to be determined dynamically, together with its gravitational effects and constraints from cosmic microwave background anisotropies. A fully consistent treatment would hence require solving the coupled scalar field and gravitational equations, or employing an appropriate thin wall description when justified. The discussion of its stability, lifetime and observational consequences would require a dedicated investigation which is beyond the scope of the current work.

\section{Implications for $H_0$ tension}

The framework developed above also allows us to ask whether a spatial dark energy deficiency could affect the value of the Hubble constant inferred from the low redshift distance ladder. The relevant observable in this case would not be the local expansion rate alone, but the luminosity distance measured by an observer who subsequently interprets the observations assuming a homogeneous cosmological background. A spatially varying dark energy component can hence generate an apparent value of $H_0$ which differs from the asymptotic cosmological value even when the global expansion history remains unchanged. We describe the spatial extent of the deficient region using the smooth profile
\begin{equation}
W_v(r)=\frac{1}{2}\left[1-\tanh\left(\frac{r-R_v}{\Delta_v}\right)\right]
\end{equation}
where $R_v$ denotes the characteristic radius of the region and $\Delta_v$ determines the width of the transition to the cosmological background. We stress that this profile represents a phenomenological description of the spatial variation in dark energy density at a chosen epoch, rather than a physical domain wall or an exact stationary solution of the underlying field equations. Its assumed time independence is an illustrative approximation and its applicability requires the spatial configuration to evolve sufficiently slowly over the observational interval considered. For the illustrative estimates below we take $\Delta_v=0.15R_v$, and we note that the precise choice of transition profile is not essential for the qualitative argument, provided that the expansion approaches the void solution near the center and the homogeneous background sufficiently far away. At low redshift we parametrize the radial expansion rate as
\begin{equation}
H_{\parallel}^{2}(r,z)=H_{\Lambda}^{2}(z)+\left[H_{v{\rm hom},0}^{2}-H_{0}^{2}-H_{0}^{2}\Omega_{{\rm DE}0}A_{\rm DE}\right]W_v(r)
\label{Hparallel_h0}
\end{equation}
The second contribution describes the enhanced expansion associated with the nonlinear matter void and its effective curvature, while the final contribution accounts for the suppression generated by the dark energy deficiency. The construction here approaches the homogeneous cosmological expansion outside the deficient region. So, at the center of the region and at the present epoch \eqref{Hparallel_h0} reduces to
\begin{equation}
H_{v0}^{2}=H_{v{\rm hom},0}^{2}-H_{0}^{2}\Omega_{{\rm DE}0}A_{\rm DE}
\label{Hcentral_h0}
\end{equation}
The matter deficiency therefore enhances the local expansion relative to the background, whereas the dark energy deficiency partially suppresses this enhancement. The resulting local value can consequently remain larger than the global $H_0$ even when the dark energy density within the same region is smaller than the cosmological mean. We first consider the limiting case in which the observer is located at the symmetry center of the approximately spherical deficient region. In this case the distance redshift relation remains isotropic and the inferred value of $H_0$ corresponds to the monopole of the more general directional expansion field. This centered construction provides the simplest estimate of the amplitude and radial scale required to affect the local distance ladder, while the consequences of relaxing the centered observer assumption will be considered below. To determine what an observer would actually infer from distance measurements, we consider the radial propagation of photons through the region. At sufficiently low redshift the comoving radial distance can be approximated through
\begin{equation}
\frac{dr}{dz}=\frac{c}{H_{\parallel}[r(z),z]}
\label{radial_h0}
\end{equation}
with $r(0)=0$ and the corresponding luminosity distance is then
\begin{equation}
D_L^v(z)=(1+z)r(z)
\label{DLvoid_h0}
\end{equation}
while the homogeneous reference prediction is given as
\begin{equation}
D_L^{\Lambda}(z)=\frac{c(1+z)}{H_0}\int_0^z\frac{dz'}{\sqrt{\Omega_{m0}(1+z')^3+\Omega_{{\rm DE}0}}}
\label{DLlcdm_h0}
\end{equation}
The quantity measured by a distance ladder analysis can then be represented by the value of $H_0$ which a homogeneous model would require in order to reproduce the inhomogeneous luminosity distance. We therefore define,
\begin{equation}
H_0^{\rm app}(z)=H_0\frac{D_L^{\Lambda}(z)}{D_L^v(z)}
\label{H0app}
\end{equation}
This is the central quantity for connecting spatial dark energy structure to the $H_0$ tension. At sufficiently small redshift $H_0^{\rm app}$ approaches the central expansion rate, whereas at distances well beyond the deficient region it approaches the asymptotic cosmological value. The same effect can be expressed directly through the distance modulus residual,
\begin{equation}
\Delta\mu(z)=5\log_{10}\left[\frac{D_L^v(z)}{D_L^{\Lambda}(z)}\right]
\end{equation}
or equivalently through
\begin{equation}
\Delta\mu(z)=-5\log_{10}\left[\frac{H_0^{\rm app}(z)}{H_0}\right]
\label{muresidual_h0}
\end{equation}
The model hence predicts not only a shift in the locally inferred value of $H_0$, but also a characteristic redshift dependence in the Hubble diagram. This provides a substantially stronger test than comparing a single local value with the cosmological background. For an observational sample with covariance matrix $C$, the covariance weighted mean residual can be written as 
\begin{equation}
\overline{\Delta\mu}_C=\frac{\mathbf{1}^{T}C^{-1}\Delta\boldsymbol{\mu}}{\mathbf{1}^{T}C^{-1}\mathbf{1}}
\end{equation}
The value that would be obtained from a homogeneous fit is consequently
\begin{equation}
H_0^{\rm fit}=H_0 10^{-\overline{\Delta\mu}_C/5}
\label{H0fit}
\end{equation}
Equation \eqref{H0fit} provides us with the quantity that can ultimately be compared directly with a local distance ladder determination using the actual supernova redshifts and covariance matrix. For a schematic numerical estimate we retain the representative parameters used throughout this work
\begin{equation*}
H_0=67.36\ {\rm km\ s^{-1}\ Mpc^{-1}} , \quad 
\Omega_{{\rm DE}0}=0.699 , \quad 
H_{v{\rm hom},0}=77.6\ {\rm km\ s^{-1}\ Mpc^{-1}}
\end{equation*}
and compare these with a representative local distance ladder determination \cite{Breuval2024}
\begin{equation}
H_0^{\rm loc}=73.17\ {\rm km\ s^{-1}\ Mpc^{-1}}
\end{equation}
If the observer and the relevant low redshift sources lie sufficiently deep inside the deficient region, \eqref{Hcentral_h0})gives the required dark energy deficiency as
\begin{equation}
A_{\rm DE}=\frac{H_{v{\rm hom},0}^{2}-\left(H_0^{\rm loc}\right)^{2}}{H_0^{2}\Omega_{{\rm DE}0}}
\end{equation}
which gives us approximately
\begin{equation}
A_{\rm DE}\simeq0.20
\label{Ade_h0}
\end{equation}
Thus the representative nonlinear matter void considered here would require a dark energy deficiency of approximately twenty percent to reduce its homogeneous dark energy expansion from $77.6\ {\rm km\ s^{-1}\ Mpc^{-1}}$ to a value close to that inferred from the local distance ladder. The radial extent of the deficient region provides an additional and independent constraint, and for that we numerically integrate \eqref{radial_h0} over the representative low redshift interval $0.0233<z<0.15$. We can then determine the value of $A_{\rm DE}$ required to produce a mean apparent value close to $73.17\ {\rm km\ s^{-1}\ Mpc^{-1}}$. For a characteristic radius $R_v=300\ {\rm Mpc}$ we obtain approximately $A_{\rm DE}\simeq0.14$. The resulting apparent Hubble constant evolves approximately from
\begin{equation}
H_0^{\rm app}(0.0233)\simeq74.65\ {\rm km\ s^{-1}\ Mpc^{-1}}
\end{equation}
to
\begin{equation}
H_0^{\rm app}(0.15)\simeq70.86\ {\rm km\ s^{-1}\ Mpc^{-1}}
\end{equation}
\begin{figure}[!h]
    \centering
    \includegraphics[width=1\linewidth]{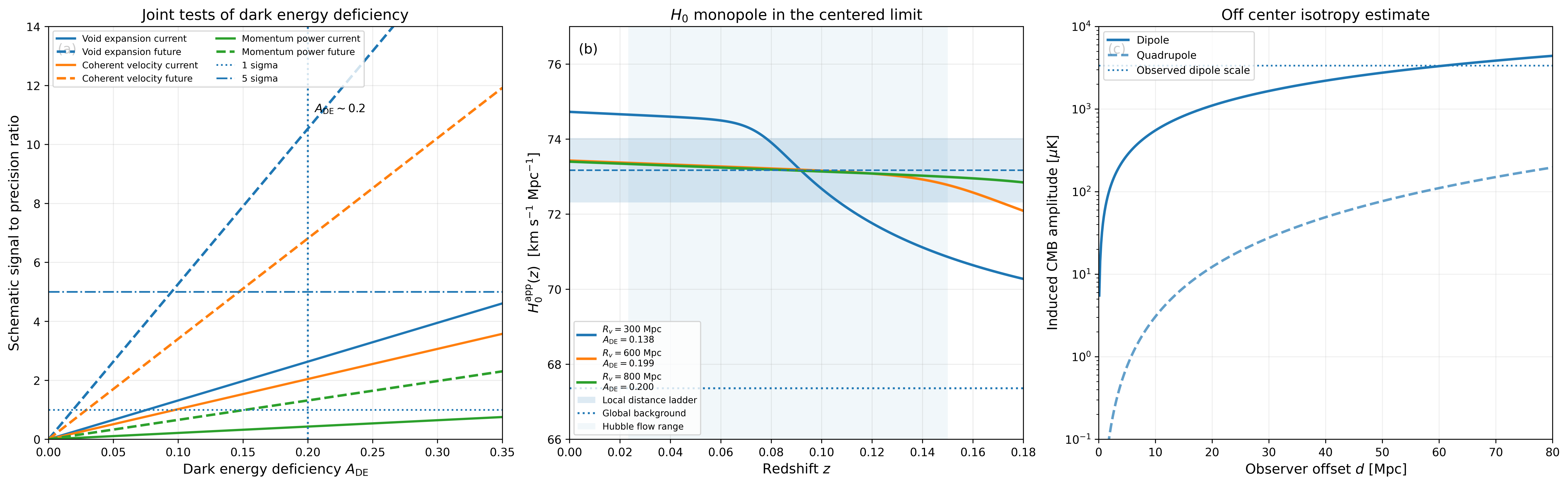}
    \caption{Joint illustration of dark energy deficiency tests, its possible impact on the apparent $H_0$ monopole and the additional isotropy constraint arising for an off center observer.}
    \label{dedefh0}
\end{figure}
A region of this size can hence reproduce the average local value but predicts a substantial redshift dependence across the Hubble flow sample. The shape of $H_0^{\rm app}(z)$ consequently provides a direct test of the spatial extent of the proposed deficiency. For a more extended region with $R_v=600\ {\rm Mpc}$, we see that the required amplitude becomes approximately $A_{\rm DE}\simeq0.20$, and the corresponding apparent values are
\begin{equation}
H_0^{\rm app}(0.0233)\simeq73.37\ {\rm km\ s^{-1}\ Mpc^{-1}}
\end{equation}
and
\begin{equation}
H_0^{\rm app}(0.15)\simeq72.78\ {\rm km\ s^{-1}\ Mpc^{-1}}
\end{equation}
The variation over the same redshift interval is therefore only
\begin{equation}
\Delta H_0^{\rm app}\simeq0.59\ {\rm km\ s^{-1}\ Mpc^{-1}}
\end{equation}
Relatively compact deficient regions can hence  reproduce the mean local value but generally predict a strong evolution of the apparent Hubble constant with redshift. More extended regions instead behave approximately as a shifted local Hubble scale throughout the low redshift distance ladder. Within the present schematic treatment, maintaining a nearly constant value close to $73\ {\rm km\ s^{-1}\ Mpc^{-1}}$ throughout the relevant Hubble flow therefore favors parameters of order $A_{\rm DE}\sim0.2$ and $R_v\gtrsim600\ {\rm Mpc}$. 
\\
\\
The large radial scale required by this centered calculation introduces an additional and independent consistency condition. An observer displaced from the symmetry center of such a region does not in general measure an isotropic distance redshift relation. Let the observer be located at displacement vector $\mathbf{d}$ from the void center, with $d=|\mathbf{d}|$, and let a source be observed in direction $\hat{\mathbf{n}}$, then defining
\begin{equation}
\mu=\hat{\mathbf{d}}\cdot\hat{\mathbf{n}}
\end{equation}
the distance of a point along the photon trajectory from the void center is
\begin{equation}
r(s,\mu)=\sqrt{d^{2}+s^{2}+2ds\mu}
\label{offcentergeometry}
\end{equation}
where $s$ denotes the distance measured from the observer along the line of sight. The low redshift propagation equation consequently becomes
\begin{equation}
\frac{ds}{dz}=\frac{c}{H_{\parallel}[r(s,\mu),z]}
\label{offcenterdistance}
\end{equation}
and the luminosity distance acquires an angular dependence
\begin{equation}
D_L^v(z,\hat{\mathbf{n}};d)=(1+z)s(z,\hat{\mathbf{n}};d)
\end{equation}
The apparent Hubble constant is therefore promoted from a function of redshift alone to
\begin{equation}
H_0^{\rm app}(z,\hat{\mathbf{n}};d)=H_0\frac{D_L^\Lambda(z)}{D_L^v(z,\hat{\mathbf{n}};d)}
\label{directionalH0}
\end{equation}
The centered result derived above is recovered in the limit $d\rightarrow0$. 
\\
\\
The directional field may be decomposed into spherical harmonics as
\begin{equation}
H_0^{\rm app}(z,\hat{\mathbf{n}};d)=\overline{H}_0^{\rm app}(z;d)+\sum_{\ell=1}^{\infty}\sum_{m=-\ell}^{\ell}a_{\ell m}^{H}(z;d)Y_{\ell m}(\hat{\mathbf{n}})
\label{H0multipoles}
\end{equation}
For a spherical region with the polar axis chosen along $\mathbf{d}$ the problem is axisymmetric and the same expansion may be written as
\begin{equation}
H_0^{\rm app}(D,\mu)=\overline{H}_0^{\rm app}(D)+H_1(D)P_1(\mu)+H_2(D)P_2(\mu)+\cdots
\end{equation}
For an observer displacement small compared with the radial scale over which $H_{\parallel}$ varies, expansion of \eqref{offcentergeometry} gives the leading dipole
\begin{equation}
H_1(D)=\frac{d}{D}\left[H_{\parallel}(D)-H_{\parallel}(0)\right]
\label{H0dipole}
\end{equation}
while the quadrupole appears at the next order as
\begin{equation}
H_2(D)=\frac{d^2}{3D}\int_0^D\left[H_{\parallel}''(s)-\frac{H_{\parallel}'(s)}{s}\right]ds
\label{H0quadrupole}
\end{equation}
Thus an off center observer generically acquires a dipolar anisotropy at first order in $d$ and a quadrupolar anisotropy at second order. The observed isotropy of the low redshift Hubble diagram therefore constrains the allowed displacement independently of the monopole requirement considered above.
\\
\\
The same geometry also produces an anisotropic imprint on the CMB. A useful schematic estimate can be obtained by approximating the region as having radius $R_v$ and an expansion contrast $\Delta H=H_{\rm in}-H_{\rm out}$. The distance from an off center observer to the spherical boundary in direction $\mu$ is
\begin{equation}
L(\mu)=-d\mu+\sqrt{R_v^2-d^2(1-\mu^2)}
\end{equation}
and for $d\ll R_v$ this becomes
\begin{equation}
L(\mu)\simeq R_v-dP_1(\mu)-\frac{d^2}{3R_v}+\frac{d^2}{3R_v}P_2(\mu)
\end{equation}
The corresponding leading geometric contribution to the CMB temperature anisotropy scales as
\begin{equation}
\left.\frac{\Delta T}{T}\right|_v\simeq\frac{\Delta H d}{c}P_1(\mu)-\frac{\Delta H d^2}{3R_vc}P_2(\mu)
\label{CMBoffcenter}
\end{equation}
so that the induced dipole is first order in the observer displacement while the quadrupole is second order. The total observed CMB multipoles must more generally be treated as
\begin{equation}
a_{\ell m}^{\rm obs}=a_{\ell m}^{\rm primordial}+a_{\ell m}^{\rm kin}+a_{\ell m}^{v}
\label{CMBmultipoles}
\end{equation}
and therefore the void contribution should ultimately be constrained through a full likelihood rather than by demanding that it individually equal the observed low multipoles. Nevertheless, \eqref{CMBoffcenter} demonstrates that the large regions required by the $H_0$ monopole calculation can impose a strong centering requirement. For the representative $R_v=600\ {\rm Mpc}$ and $A_{\rm DE}\simeq0.20$ configuration, the central expansion differs from the asymptotic background by approximately
\begin{equation}
\Delta H\simeq5.8\ {\rm km\ s^{-1}\ Mpc^{-1}}
\end{equation}
An illustrative displacement of $d=20 {\rm Mpc}$ would then give a geometrically induced CMB dipole amplitude of order
\begin{equation}
T_{\rm CMB}\frac{\Delta H d}{c}\sim10^3\ \mu{\rm K}
\end{equation}
while the corresponding quadrupole is of order a few $\mu{\rm K}$. These numbers are not a substitute for an exact CMB calculation, but they show that observer position cannot be neglected when assessing the viability of the $H_0$ interpretation. In particular, the total observed CMB dipole shown in Fig. \ref{dedefh0} should be regarded only as a visual reference since most of the measured dipole is already associated with the observer's peculiar motion and the admissible additional contribution from the deficient region must be smaller.
\\
\\
These values should not be interpreted as a fit to the $H_0$ tension and instead they should be thought to establish the approximate amplitude and spatial extent that such an environmental explanation would require within the present phenomenological framework. The large characteristic scale itself provides an important observational constraint on the scenario too. The parameter space relevant for the single spherical region interpretation is therefore enlarged from $\{A_{\rm DE},R_v\}$ to at least
\begin{equation}
\left\{A_{\rm DE},R_v,d,\hat{\mathbf{d}}\right\}
\end{equation}
and a viable realization must reproduce the required monopole in $H_0^{\rm app}$ while simultaneously satisfying the directional isotropy of the low redshift distance ladder and the observed CMB multipoles. A configuration that produces the correct local $H_0$ only for an observer confined to an extremely small central volume would represent a strongly fine tuned realization of the scenario. Conversely, the survival of these anisotropy constraints would provide an additional nontrivial prediction rather than merely reproducing the monopole expansion rate. We reach to a very interesting conclusion now, which is that the $H_0$ tension supplies a useful closing consistency test for the framework developed in this work. The values of $A_{\rm DE}$ and $R_v$ capable of shifting the luminosity distance cannot be chosen independently of the void dynamics. The same spatial configuration must reproduce the nonlinear local expansion, the coherent peculiar velocity field, the density and momentum power spectra and the luminosity distance residual. For a single approximately spherical region, it must additionally reproduce the observed near isotropy of the Hubble diagram and remain compatible with the CMB dipole, quadrupole and higher angular structure for the same observer location. A model which reproduces the inferred local value of $H_0$ but fails these independent kinematic tests would consequently be excluded.
\\
\\
Figure 1 summarizes the resulting three stage consistency test, providing a complete view of this scenario. The left panel shows the schematic sensitivity of void expansion, coherent peculiar velocity and momentum power measurements to increasing dark energy deficiency, with the $A_{\rm DE}\sim0.2$ regime required by the illustrative $H_0$ application lying within the range that future kinematic measurements could probe. The middle panel shows the centered monopole prediction for $H_0^{\rm app}(z)$, where compact regions produce substantial redshift evolution while $R_v\sim600$-$800$ Mpc configurations maintain an approximately constant value near $73\ {\rm km\ s^{-1}\ Mpc^{-1}}$ across the low redshift Hubble flow. The right panel then illustrates the additional off center consistency condition for the representative extended configuration, showing the rapid growth of the induced CMB dipole and quadrupole with observer displacement. The horizontal CMB dipole line is the total observed dipole scale and is shown only as a reference rather than as the permitted contribution from the deficient region. A successful explanation of the $H_0$ tension must hence satisfy the kinematic, monopole and isotropy tests simultaneously with the same spatial configuration.

\section{Where to find DE deficient regions?}

The preceding analysis provides not only a set of observables for identifying dark energy deficient regions, but also some guidance concerning where such regions should be searched for most prominently. If the same phenomenon is to contribute significantly to the $H_0$ tension, then the preferred targets are not arbitrary cosmic voids. The most interesting environments are the largest, deepest and lowest redshift underdensities whose spatial extent is sufficiently large to influence a significant fraction of the local distance ladder. For the representative nonlinear void considered above, the $H_0$ analysis indicated characteristic requirements of approximately $A_{\rm DE}\sim0.2$ and $R_{\rm DE}\gtrsim600\ {\rm Mpc}$ in order to maintain an apparent value close to the locally inferred $H_0$ across much of the low redshift Hubble flow. For a single approximately spherical region, these requirements must be supplemented by the observer position constraint derived above. Candidate structures relevant to the $H_0$ explanation should hence be assessed not only according to their depth and spatial extent but also according to whether their inferred center and geometry permit the observer location to remain consistent with the observed isotropy of the CMB and the low redshift Hubble diagram. These numbers are model dependent, but they provide a useful scale against which known cosmic underdensities can be assessed. In particular, they suggest that isolated small voids are unlikely to provide the dominant effect if the $H_0$ tension is involved, while very extended void complexes and coherent underdense environments become especially interesting.
\\
\\
An immediate target would then be the large local underdensity commonly associated with the Keenan- Barger-Cowie measurements \cite{keenan2013evidence}. Galaxy luminosity density measurements have indicated an underdense local environment extending over scales of order several hundred Mpc, with the luminosity density increasing toward larger distances. Since the observer and the nearby distance ladder are located within this environment, it provides the most direct laboratory for testing whether the locally inferred expansion contains a component that is not explained by the measured matter distribution alone. Its relevance to a single region explanation of the $H_0$ tension would, however, also require reconstruction of its center, morphology and our displacement relative to it. The same reconstruction that determines the radial matter and expansion profiles could therefore be used to predict the directional multipoles in $H_0^{\rm app}$ and the associated CMB anisotropy. We emphasize that the existence, depth and cosmological significance of such a local underdensity remains debated, and supernova analyses place strong constraints on conventional matter only void explanations of the $H_0$ tension \cite{kenworthy2019local}. This makes it particularly useful as a falsifiable test of the present framework rather than evidence for the framework itself. An important possibility is that the characteristic radius of a dark energy deficiency need not coincide with the radius inferred from the galaxy or matter distribution. In general one may have \begin{equation}
R_{\rm DE}\neq R_m
\end{equation}
and in particular a scenario with
\begin{equation}
R_{\rm DE}>R_m
\end{equation}
would be especially interesting in view of the large scale suggested by the $H_0$ calculation. The matter density could return approximately to its cosmic mean while a dark energy induced expansion residual persists to larger distances. Such a configuration would be difficult to identify using galaxy counts alone and motivates the joint use of peculiar velocities, distance indicators and gravitational potential measurements. A particularly sharp signature would hence occur if observations find a radial regime satisfying schematically
\begin{equation}
\delta_m(r)\simeq0 \qquad \Delta H_v(r)\neq0
\end{equation}
after the ordinary nonlinear matter contribution has been reconstructed. In such a region the measured matter profile would no longer provide an obvious explanation for the residual expansion. The radial extent of the kinematic anomaly could then be compared directly with the matter boundary and with the $W_v(r)$ profile inferred for the dark energy deficiency.
\\
\\
A second particularly interesting target is the Eridanus supervoid associated with the direction of the CMB Cold Spot \cite{kovacs2016cosmic}. Reconstructions here indicate that a central matter underdensity of order $\delta_m\sim-0.25$ and an unusually extended line of sight structure, with a reported characteristic extent approaching several hundred Mpc. Its combination of large physical size, substantial matter deficiency and existing CMB information makes it well suited to a joint test of the two classes of signatures developed in this work. Peculiar velocities and the local density field can test the expansion and momentum sector, while lensing and the integrated Sachs-Wolfe signal can independently probe the evolution of the gravitational potential. The Eridanus direction also provides a useful opportunity for environmental comparison, as its antipodal line of sight intersects the Northern Local Supervoid at low redshift and subsequently several prominent overdense structures including the Hercules and Corona Borealis superclusters and the regions associated with the Coma and Sloan Great Walls \cite{kovacs2016cosmic}. Such underdense and overdense environments can be compared using the same estimators. A genuine spatial dark energy effect associated preferentially with voids should correlate with environment rather than appearing as an approximately universal velocity or distance calibration offset. Because Eridanus occupies a preferred direction on the sky rather than surrounding the observer approximately isotropically, it is particularly useful as an independent laboratory for dark energy deficiency but should not by itself be identified with the single spherical structure required to generate an isotropic local $H_0$ monopole. Its CMB alignment instead makes it valuable for testing whether extreme underdense environments exhibit correlated kinematic and gravitational potential anomalies.
\\
\\
This suggests a general observational strategy in which clusters and superclusters are used primarily as control environments rather than as the preferred locations in which to search for a deficiency. One may compare an environment conditioned apparent Hubble parameter through
\begin{equation}
\Delta H_0^{\rm env}(z)=H_0^{\rm void}(z)-H_0^{\rm control}(z)
\end{equation}
where the control sample consists of sightlines or sources at comparable redshift but located in average density or overdense environments. A correlation between the inferred expansion residual and the integrated void environment would provide information that cannot be obtained from the sky averaged value of $H_0$ alone. For sufficiently large candidate regions this comparison should also retain the directional information rather than compressing the data immediately into a sky averaged quantity. The environment dependent monopole, dipole and quadrupole of the distance residual provide complementary information, as the monopole tests the possible contribution to the $H_0$ tension while the higher multipoles constrain observer displacement and departures from isotropy. Large statistical void catalogues offer a complementary approach, as DESI DR1 already provides hundreds to more than one thousand low redshift voids depending on the void finder employed \cite{qin2026desi}. Most individual objects are far smaller than the several hundred Mpc scale suggested by the illustrative $H_0$ calculation, but collectively they provide an important population test. The framework predicts that any environmental expansion residual should be studied as a function of void radius, depth, morphology and isolation rather than simply averaged over all identified voids.
\\
\\
One can therefore search for an empirical trend of the form
\begin{equation}
\Delta H_v=\Delta H_v(R_v,\delta_m,A_{\rm DE})
\end{equation}
and compare isolated deep voids with shallower voids, interacting voids and matched overdense regions. Such a population analysis would also help distinguish genuine dark energy related behavior from peculiar velocity systematics or inaccuracies in the nonlinear void reconstruction. A distributed population also provides a possible alternative to a single exceptionally large spherical region. Multiple deficient environments could contribute to the low redshift distance field while their directional signatures partly cancel, although such a realization would require a separate statistical treatment and cannot be assumed to evade the isotropy constraint automatically. The most informative targets can consequently be divided into three classes. The local underdense environment is the highest priority for testing the possible connection to the $H_0$ tension because the observer and the nearby distance ladder are embedded within it. Extremely extended structures such as the Eridanus supervoid provide independent laboratories in which expansion, velocity, lensing and ISW signatures can be tested together. Large catalogues such as those from DESI would then provide the statistical control needed to determine whether any inferred anomaly systematically tracks cosmic environment. A particularly compelling outcome would be the observation of a kinematic or distance residual whose spatial extent exceeds that of the reconstructed matter underdensity and which appears preferentially in the largest and deepest void environments. For the specific interpretation in which a single local deficient region accounts for the $H_0$ tension, such a detection would additionally have to give an observer location compatible with the observed small angular anisotropy. Conversely, the absence of such residuals in these specifically selected systems would place direct limits on $A_{\rm DE}$ and would strongly constrain the possibility that spatial dark energy structure contributes appreciably to the $H_0$ tension. The framework would hence identify not only how dark energy deficient regions could be detected, but also where the most decisive searches for them should be performed.
\\
\\
For the sake of completeness, we also mention that one can use simulations of cosmological void catalogues as an important complementary approach to identifying possible dark energy deficient regions in the near future. In particular, cosmological $N$-body and hydrodynamical simulations, such as IllustrisTNG \cite{tng1,tng2,tng3,tng4,tng5} can provide realistic populations of voids with different radii, density contrasts, compensating walls and environmental morphologies, allowing the nonlinear expansion framework developed here to be tested against a homogeneous dark energy null hypothesis. Such an exercise would be especially valuable because ordinary matter voids can themselves generate enhanced local expansion rates and apparent deviations in the inferred Hubble parameter, potentially mimicking the signatures of a spatial dark energy deficiency if their matter distributions and peculiar velocity fields are not reconstructed consistently. An important requirement would hence be to account for the compensating overdense structures surrounding realistic voids, rather than relying exclusively on simplified radial expansion profiles and to calibrate the expected relation between the enclosed matter contrast, coherent peculiar velocities and local expansion rate. Mock observers placed at different positions within and around simulated voids could further be used to investigate the resulting directional dependence of $H_0^{\mathrm{app}}$, including its monopole, dipole and quadrupole contributions, and to quantify the possibility of false positive inferences of $A_{\mathrm{DE}}>0$ in an otherwise homogeneous dark energy cosmology. One could then compare these calibrated null predictions with simulations incorporating clustering or environmentally coupled dark energy, thereby determining whether genuine spatial deficiencies produce distinguishable signatures in the expansion and velocity fields. Applying such an analysis to the observational void catalogues from DESI and future large scale structure surveys would provide a particularly useful consistency test of the framework proposed here, but discussing them in detail here lies beyond the scope of the current work.

\section{Conclusions}
So, what is the big claim one can most seriously make from this work? The $H_0$ tension is an observed discrepancy between the expansion rate inferred from the local distance ladder and that inferred from the early Universe under the standard cosmological model. If spatially dark energy deficient regions are indeed responsible for a significant part of this discrepancy, then the framework we have here makes a concrete and falsifiable prediction: the largest nearby underdense environments should exhibit an expansion residual that cannot be accounted for by their measured matter deficiency alone. The most compelling targets are hence the local underdense environment surrounding the observer, exceptionally extended structures such as the Eridanus supervoid and the largest and deepest void systems identified in surveys such as DESI. In these regions, the tests proposed here should be applied jointly through nonlinear void expansion, peculiar velocities, density and momentum statistics, supernova distance residuals, lensing and gravitational potential evolution. For a single approximately spherical region to explain the local $H_0$ monopole, the claim is more restrictive. In that case, the same spatial configuration must reproduce the required isotropic shift in the distance scale without generating an excessive directional Hubble anisotropy or CMB dipole and quadrupole for the actual observer location. The observer displacement would itself be a falsifiable parameter of the model rather than an arbitrary geometrical choice.
\\
\\
Our most ambitious prediction would be to consequently say that such searches will identify coherent regions extending over several hundred Mpc in which the inferred kinematics are consistent with a genuine dark energy deficiency, with the characteristic amplitude potentially reaching the tens of percent level required by the $H_0$ application considered here. If the full $H_0$ discrepancy is produced by one such spherical region, the same reconstruction should additionally locate us sufficiently close to its effective center that the predicted angular multipoles remain compatible with the observed isotropy. Failure of this centering and isotropy condition would strongly disfavor the single region explanation even if dark energy deficient environments themselves were independently detected. In particular, if the same spatial profile that produces an apparent $H_0$ near the locally measured value also predicts the observed peculiar velocity residuals, momentum statistics and potential evolution without independent retuning, the case for a spatially nonuniform dark energy component would become considerably stronger. An especially compelling result would be for this same profile to satisfy the CMB and low redshift isotropy constraints without requiring an implausibly fine tuned observer location. Conversely, failure to find such correlated signatures in the very environments where this mechanism requires them would directly constrain, and potentially rule out, dark energy deficiency as an explanation of the $H_0$ tension. The central claim of this work would be  not to merely say that dark energy deficient regions may exist, but that if they are responsible for the local expansion anomaly, their characteristic scale, amplitude and observational fingerprints are sufficiently specific that present and forthcoming surveys should be able to find them or decisively exclude them. The $H_0$ monopole, its directional multipoles and the CMB anisotropy therefore become three aspects of the same consistency test rather than independent adjustable observables.
\\
\\
The central result from our work for the $H_0$ perspective would hence be the following. If DE deficient regions are responsible for the $H_0$ tension, astronomers should find a several hundred Mpc nearby environment with an order tens of percent DE deficit whose matter profile, peculiar velocity field, momentum statistics, supernova distances, and angular isotropy all agree with one common spatial solution. That is quite a strong claim because failure at any one of those stages can kill the $H_0$ interpretation even if DE deficient regions themselves still exist.

\section*{Acknowledgments}
The authors would like to thank Robert J. Scherrer and David Benisty for very helpful discussions and comments during the writing of this work. OT is supported in part by the Vanderbilt Discovery Doctoral Fellowship. AL is supported in part by the Black Hole Initiative, which is funded by GBMF and JTF.
\bibliography{references}
\bibliographystyle{unsrt}

\end{document}